# A nonlinear magnonic nano-ring resonator

Qi Wang[1,5,*,†], Abbass Hamadeh[1,2,*], Roman Verba[3,*], Vitaliy Lomakin[4], Morteza Mohseni[1], Burkard Hillebrands[1], Andrii V. Chumak[5], Philipp Pirro[1]

[1] *Fachbereich Physik, Technische Universitaet Kaiserslautern, Kaiserslautern, Germany*
[2] *Multi-Disciplinary Physics Laboratory, Faculty of Sciences, Lebanese University, Beirut, Lebanon.*
[3] *Institute of Magnetism, Kyiv 03142, Ukraine*
[4] *Center for Magnetic Research, University of California at San Diego, La Jolla, CA, USA*
[5] *Faculty of Physics, University of Vienna, Boltzmanngasse 5, A-1090 Vienna, Austria*

The field of magnonics, which aims at using spin waves as carriers in data processing devices, has attracted increasing interest in recent years. We present and study micromagnetically a nonlinear nanoscale magnonic ring resonator device for enabling implementations of magnonic logic gates and neuromorphic magnonic circuits. In the linear regime, this device efficiently suppresses spin-wave transmission using the phenomenon of critical resonant coupling, thus exhibiting the behavior of a notch filter. By increasing the spin-wave input power, the resonance frequency is shifted leading to transmission curves, depending on the frequency, reminiscent of the activation functions of neurons or showing the characteristics of a power limiter. An analytical theory is developed to describe the transmission curve of magnonic ring resonators in the linear and nonlinear regimes and validated by a comprehensive micromagnetic study. The proposed magnonic ring resonator provides a multi-functional nonlinear building block for unconventional magnonic circuits.

---

[*] These authors have contributed equally to this work.
[†] Corresponding author: qi.wang@univie.ac.at

# Introduction

Spin waves are collective excitations of spin systems in magnetic materials, which can be considered as a potential data carrier in future low-energy data processing systems [1-4]. This is due to their small wavelengths, down to a few nanometers [5-6], high frequencies up to a few terahertz [7], ultralow losses [8-9], and abundance of associated nonlinear phenomena [10-12]. These features make spin waves highly attractive for wave-based and neuromorphic computing concepts. Several important milestones were achieved in the realization of magnonic data processing units, including logic gates [13-15], majority gates [16-18], a magnon transistor [19], a phase shifter [20], building blocks for unconventional computing [21-23], auxiliary units for integrated circuits [12], magnonic directional couplers [24-27], and an integrated magnonic half-adder [25].

Here, we propose a nanoscale nonlinear magnonic ring resonator. It is magnetic counterpart of the photonic ring resonator, which is considered as a universal unit and widely used in integrated photonic circuits [28], photonic quantum computing [29], and photonic neuromorphic computing [30]. The concept of the magnonic ring resonator (see Fig. 1(a)) is similar to that of the photonic ring resonator [31] except that spin waves, instead of light, are used to carry information. A magnonic ring resonator of submillimeter size has been studied in the linear regime using micromagnetic simulations as reported in Ref. [32]. Although such rather macroscopic ring resonators demonstrate certain interesting features due to multimode coupling and external field sensitivity, their functionality and size are hardly compatible with the current state of the CMOS technology. Moreover, the presence of multiple modes such as width modes with different coupling strengths results in less effective energy transfer and makes it impossible for the device to operate in the so-called "critical coupling" condition. Here, we study the single-mode nanoscale magnonic ring resonator using the critical coupling phenomenon and demonstrate its functionality analytically and by simulation, including linear and nonlinear operation regimes as well as their anticipated applications. Despite the fact that this is a simulation, the recent progress in the realization of single-mode magnonic nano-conduits proves that the sizes chosen here can be realized based on the current nano-fabricating technology [33].

**Theory and micromagnetic simulations of the linear regime**

The basic configuration of the magnonic ring resonator consists of a ring of mean radius $R$ and width $w$, and a straight waveguide of same width $w$, as shown in Fig. 1a. The static magnetization distribution of the magnonic ring resonator obtained from micromagnetic simulation is shown in Fig. 1b (details of the micromagnetic simulation method are described in section Methods). For a sufficiently narrow ring, here $w = 100$ nm, the static magnetization is in the vortex state with the magnetization lying along the ring. Such a vortex state is the ground state in the presence of zero external fields (i.e., it corresponds to the global energy minimum), and it can be easily achieved in experiments [34,35], for instance, by controlling the variation of the external field (see section Supplementary Materials). The static magnetization of a straight waveguide is uniform and is along the waveguide, which is caused by strong shape anisotropy and by the small cross-section. We consider Yttrium-Iron-Garnet (YIG) as the material of both the waveguide and the ring: It is chosen for its low damping allowing for long-range spin-wave propagation [36]. The used material parameters of YIG are described in section Methods.

For the theoretical description of the power transmission in the ring resonator, we adopt a method typically used in optics and microwave electronics [31, 37]. Let us denote the complex amplitudes of input (output) spin waves in the waveguide and ring by $a_{1,2}$ ($b_{1,2}$), as shown in Fig. 1a. To define a reference plane, we use the position of the minimum distance between the ring and waveguide (see the dashed line in Fig. 1a), so that all $a_i$ and $b_i$ are the values, which would be at the point $x = 0$ if continuously extrapolated in the absence of coupling. If the coupling between the ring and waveguide is lossless, which is the case for pure dipolar coupling, it is described by the unitary scattering matrix [31]:

$$\begin{pmatrix} b_1 \\ b_2 \end{pmatrix} = \begin{pmatrix} \tau & i\kappa \\ i\kappa & \tau \end{pmatrix} \begin{pmatrix} a_1 \\ a_2 \end{pmatrix} \quad (1)$$

The parameter $\kappa$ is the coupling coefficient between the straight waveguide and the ring that shows the fraction of the spin-wave amplitude coupled from the waveguide into the ring structure and vice versa. The parameter $\tau$ is the transmission coefficient across the coupling region, which shows the fraction of the spin-wave amplitude passed through the coupling region. The transmission coefficient $\tau$ is different from the transmittance $T$, which demonstrates the final transmitted power through all the structures and accounts for the interference in the ring. Naturally, $|\kappa|^2 + |\tau|^2 = 1$, which reflects the lossless nature of the coupling. The calculation of the coefficients $\kappa$ and $\tau$ for the dipolar coupled waveguides and ring is presented in section Methods.

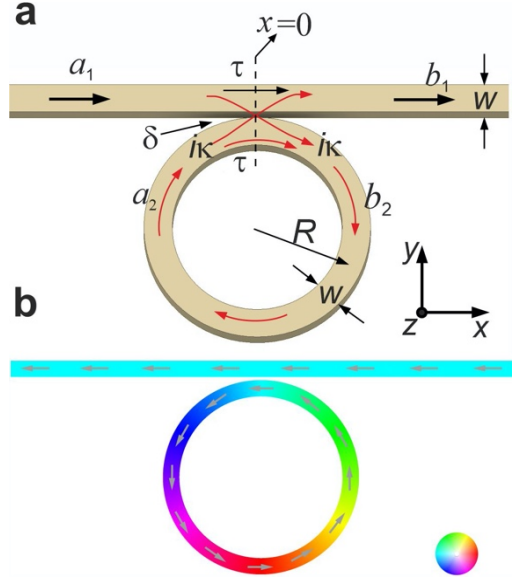

*Fig. 1 Geometry and parameters used in the modeling, and characteristics of the magnonic ring resonator. **a** Generic model of the magnonic ring resonator. The width of the waveguides is w = 100 nm, and the thickness is h = 50 nm, the radius (from the center of the ring to the center of the waveguide) is R = 550 nm, the minimum gap between the ring and waveguide is δ = 20 nm. The dashed line shows the position of the reference plane. **b** Initial magnetization distribution of the investigated structure. The grey arrows and colors represent the in-plane orientations of the magnetization **M**.*

The complex amplitudes $a_2$ and $b_2$ are connected by the circulation condition: $a_2 = b_2 \beta e^{i\theta}$. Here, $\beta = \exp(-2\pi R \Gamma / v_{gr})$ is the loss coefficient that describes which part of the spin-wave amplitude remains in the ring after one circulation, with $\Gamma$ and $v_{gr}$ being the spin-wave damping rate and group velocity in the ring, respectively (for details see section Methods). The parameter $\theta = 2\pi R k$ is the round-trip phase accumulation, where $k$ is the wavenumber of the spin wave in the ring. The wavenumber is determined by the input spin-wave frequency, which is given by the dispersion relation $\omega_k$ in the ring and is normally different from the wavenumber in the straight waveguide.

Solving Eq. (1) together with the circulation condition one finds the transmission $T$ through the ring resonator structure:

$$T = \frac{|b_1|^2}{|a_1|^2} = \frac{\beta^2 + |\tau|^2 - 2\beta|\tau|\cos(\theta - \psi)}{1 + \beta^2|\tau|^2 - 2\beta|\tau|\cos(\theta - \psi)}, \qquad (2)$$

using the phase $\psi = \text{Arg}[\tau]$. In our case of a straight waveguide and ring of the same cross-section, the phase is $\psi = 0$. At the resonance frequencies, at which $\theta = 2\pi n$ ($n = 0, 1, 2, 3, \ldots$ is the number of the resonance modes in the ring), the transmission $T$ is equal to $T = (\beta - |\tau|)^2 / (1 - \beta|\tau|)^2$. The output signal vanishes completely if the transmission coefficient is equal to the loss coefficient, i.e. $|\tau| = \beta$, which is the so-called case of "critical coupling". The maximum transmission $T$ in the critical case, which is reached when $\theta = (2n+1)\pi$, is equal to $T = 4\beta^2 / (1 + \beta^2)^2$, and increases with $\beta$. Therefore, it is desirable to work in the range of $\beta \approx |\tau| \to 1$ to achieve a large output power, i.e. to have small losses in the ring and weak coupling between the ring and the waveguide. However, a small loss and a small coupling increase the operational time of the resonator (time to reach the dynamic equilibrium) [38]. Thus, optimal values of the transmission coefficient and the loss coefficient are in the range $\beta \approx |\tau| \sim 0.6 - 0.9$, which is the result of the trade-off between a large transmission and a short delay time for the ring resonator.

The coefficients $\kappa$, $\theta$ and $\beta$ depend on the spin-wave frequency. The coupling coefficient $\kappa$ significantly depends on the gap between waveguide and ring. In our example simulations, the minimum gap, which is the closest distance between the waveguide and the ring, is fixed to $\delta = 20$ nm for all simulations. The ring radius determines the separation between the ring resonance frequencies. To set the loss coefficient to a value close to the optimum of $\beta \approx |\tau|$, in our simulations we increased the Gilbert damping in the ring structure to $\alpha_G = 2 \times 10^{-3}$. In an experiment, an increase of the YIG damping can be realized, for instance, by placing a normal metal on top of the ring to use the phenomenon of spin pumping [39]. Please note that the parameters $R$, $\delta$ and $\alpha_G$ are selected for the working frequency range from 2.6 GHz to 2.8 GHz. These parameters can also be modified to obtain other frequency ranges that fulfill the critical coupling condition.

As an approximation of the ring dispersion relation, in principle, one can use the dispersion relation of a straight waveguide [40]. For our case it results in only a slight discrepancy of 80 MHz, and the discrepancy becomes more negligible for $R \gg w$ and $kR \gg 1$. In all the following calculations, we use a more accurate theory of the dispersion in the ring, as outlined in section Methods. The spin-wave damping rate and group velocity, which determine the loss coefficient, are calculated from the dispersion relation.

The frequency dependencies of the transmission and the loss coefficient $\tau$ and $\beta$, together with the round-trip phase $\theta$ are shown in Fig. 2. In the chosen frequency range the condition $\tau \sim \beta$ holds, and the critical coupling condition is exactly satisfied at a frequency around 2.55 GHz (not shown in Fig. 2). The frequency dependence of the transmission coefficient is more pronounced because of a significant wavenumber dependence of the dynamic dipolar fields, generated by the spin-waves propagating in the waveguide and the ring [24].

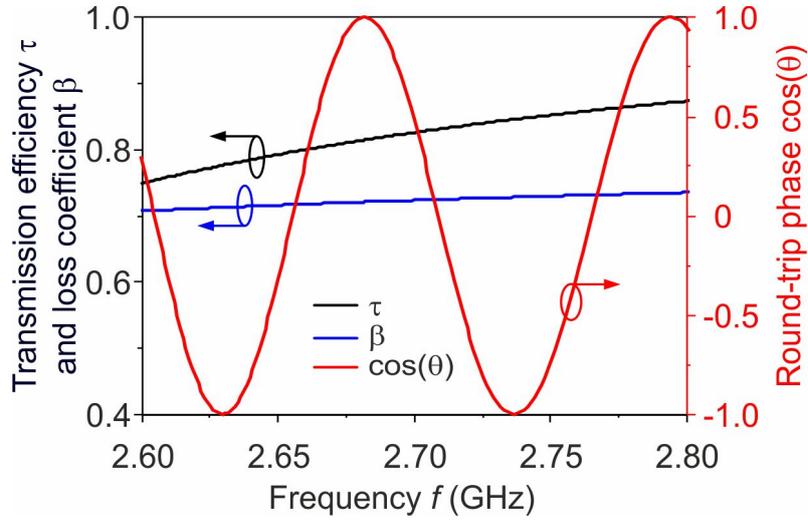

*Fig. 2 Transmission coefficient ($\tau$), loss coefficient ($\beta$), and round-trip phase cos($\theta$) as a function of frequency f.*

The theoretical transmission curve of the whole ring resonator calculated according to Eq. (2), as well as the results of micromagnetic simulations are shown in Fig. 3a. In the simulation, the transmission $T$ is defined by calculating the ratio of the spin-wave intensities in the straight waveguide behind and in front of the ring. Two resonance frequencies of the magnonic ring resonator are observed in this frequency range, which correspond to the 16[th] and 17[th] resonant mode. At these frequencies, the output signal is vanishing due to the destructive interference in the outgoing waveguide between the transmitted spin wave $a_1\tau$ and the coupled-back spin wave $ia_2\kappa$, which acquires the round trip phase of $\theta = 2\pi n$ plus two $\pi/2$ phase shifts in the coupler (see Eq. (1)), being, in total, in antiphase to the first wave. At the resonance frequency, all the spin-wave power is concentrated in the ring (see Fig. 3b). In contrast, at the frequency of 2.724 GHz, which corresponds to $\theta = 2\pi \times 16.5$, the constructive interference conditions are satisfied, and a large part of input power is transmitted, while only a small amount is circulating in the ring (Fig. 3b). A strong frequency dependence of the output power is important as it allows one to realize notch filters with a magnonic ring, and it enhances the sensitivity of the system to a nonlinear frequency shift.

In addition, Fig. 3 shows the transmission curves for the rings having different radii. As expected, the resonance frequencies change and the resonance curves are shifted, while preserving their shape. The resonance frequencies, calculated analytically, are 20 MHz higher than those found from the micromagnetic simulations. Also, in the simulations, the critical coupling condition is satisfied in the range 2.65 GHz-2.68 GHz, as it is evident from the vanishing output at the resonance frequency, while theory indicates the critical coupling at a slightly different frequency of 2.55 GHz. These two weak discrepancies are mainly due to a slightly nonuniform width profile of the spin waves in the ring and the waveguide, and to a weak dipolar field generated by the straight waveguide, which slightly modifies the dispersion relation in the ring. Both effects are not taken into account in the theory. Furthermore, there is a certain difference in the maximum transmission energy between the simulations and the theoretical calculations, which is attributed to the propagation losses in the straight waveguide and the coupling area.

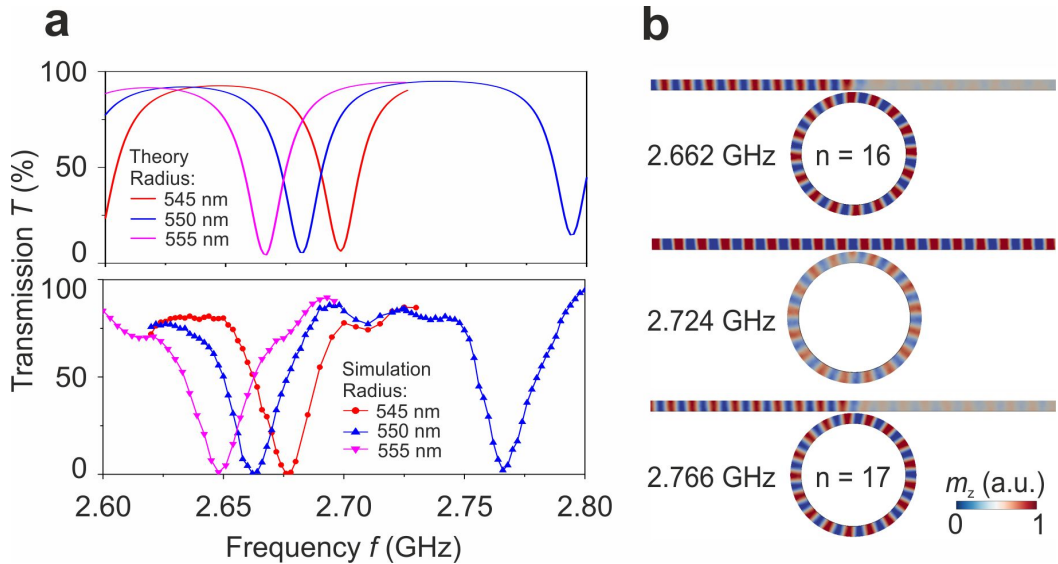

*Fig. 3 **a** Theoretical (top panel) and simulated (bottom panel) normalized transmission T as a function of frequency f for different radii of the ring. **b** Snapshot of the out-of-plane component spin-wave amplitude $m_z$ in the ring resonator with the radius of 550 nm for different excitation frequencies.*

### Reconfigurable magnonic ring resonator

In the previous section, the magnetization in the ring is oriented counterclockwise and the magnetization in the coupling region between ring and waveguide is aligned parallel. However, in the absence of an external field, the ring can exist in two stable magnetic configurations – clockwise and counterclockwise - as shown in the insets of Fig. 4. These two states lead to antiparallel and parallel magnetization configurations in the coupling region. The switching between these two states can be realized by controlling the tracks of the variation of the external field before reaching the remanent state (for details see section Supplementary Materials). The coupling strength, described by the coupling coefficient κ, depends strongly on the static magnetization configuration and affects significantly the transmission coefficient τ [24]. The coupling strength is stronger for the antiparallel magnetization configuration and this results in the breaking of the critical coupling condition, i.e., $|\tau| \neq \beta$ and consequently, the transmission $T$ at the resonance frequencies increases. Figure 4 shows the transmission curve for parallel and antiparallel alignments in which the transmissions at the resonance frequency of 2.662 GHz are 0.48% and 30.8%, respectively. The contrast between the two states can be further increased by optimizing the parameters of the ring resonator. For a future on-chip magnonic device, this switching can be realized by a local Oersted field created by direct current passing through a conducting wire which is placed on top of the ring structure. This example shows that the symmetry break caused by the direction of the magnetization allows creating magnonic functionalities that are not available in the same form in, e.g., photonics.

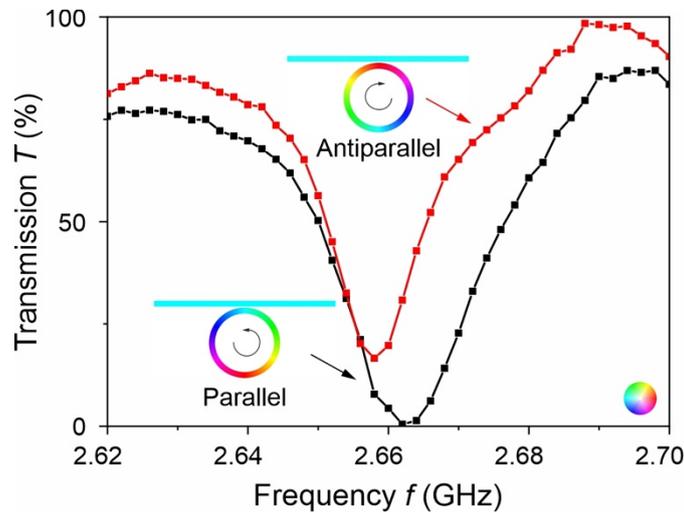

*Fig. 4 Simulated transmission T as a function of frequency f for different initial magnetization orientations. The insets show the initial magnetic configurations. The color-disk represents the in-plane orientations of the magnetization **M***

# Nonlinearity of magnonic rings

Magnonic systems are known to involve a variety of nonlinear effects, which open a way for the development of various nonlinear power-dependent devices. In general, all the parameters, which define the operation of the magnonic ring resonator, namely $\theta$, $\beta$ and $\tau$, are power-dependent. However, it can be shown that the main impact of nonlinearities is caused by the nonlinear phase accumulation $\theta = \theta(b)$, where $b$ is the spin-wave amplitude, while the nonlinearities of the loss coefficient (due to the group velocity shift) and the coupling strength lead only to a small (second-order) correction, and, therefore, can be neglected in almost all experimentally achievable cases.

An increase of the spin-wave power results in a nonlinear shift of the spin-wave resonance frequency, $\omega_k(b) = \omega_k^{(lin)} + W_k |b|^2$, where $W_k$ is the nonlinear shift coefficient. Consequently, a wave of a constant frequency possesses a power-dependent wavenumber $k \approx k_0 - (W_k / v_{gr}) |b|^2$, which directly affects the phase accumulation during the spin-wave propagation. The integration over the ring yields the round-trip phase $\theta(b_2) = 2\pi R(k_0 - K|b_2|^2)$, where $K = W_k (1 - e^{-4\pi\Gamma R/v_{gr}}) / (4\pi\Gamma R) \approx W_k \beta / v_{gr}$ is the averaged coefficient of the nonlinear shift of the spin-wave wavenumber. Then, Eq. (1) together with the circulation condition yields the following relation:

$$|b_2|^2 = |a_1|^2 \frac{\kappa^2}{1 + \beta^2 \tau^2 - 2\beta\tau \cos[\theta(b_2) - \psi]}, \quad (3)$$

which implicitly determines the amplitude in the ring. The transmission $T$ is given by the same Eq. (2), in which one should use the nonlinear phase accumulation $\theta(b_2)$ with the amplitude of $b_2$, found from Eq. (3).

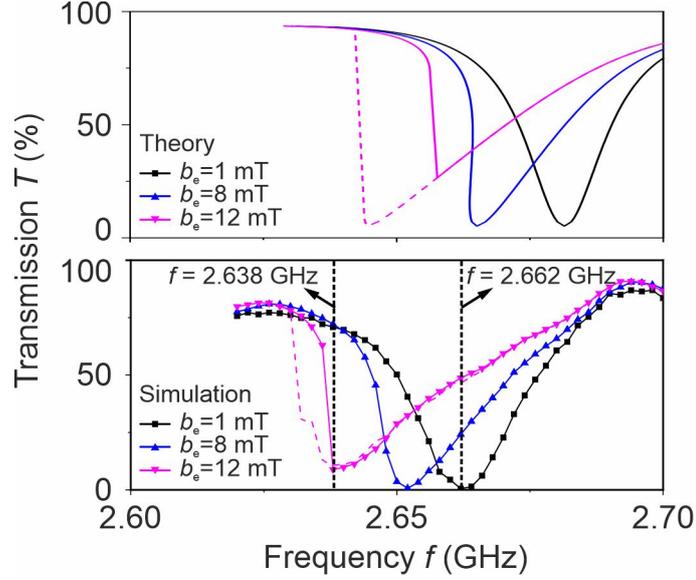

*Fig. 5 Transmission as a function of spin-wave frequency for different values of the excitation field $b_e$ (top panel: theory, bottom panel: simulation). Dashed magenta lines represent the second solution in the bistability range for $b_e = 12$ mT.*

The simulated transmission curves of the magnonic ring resonator for different excitation fields $b_e$ are shown in Fig. 5 (bottom panel). A pronounced shift of the resonance frequency, at which transmission is minimum, is observed. A similar power-dependent transmission curve shift was observed in optical ring resonators [28, 41, 42]. To plot the theoretical curves, we use a nonlinear frequency shift value of $W = -2\pi \times 2.6$ GHz, which is calculated for a straight waveguide [25, 43]. As one can see, this approximation is reasonable and gives a similar shift of the transmission minimum position compared to the simulated results.

For a large enough input spin-wave power, the transmission curve becomes bistable (see the magenta line in Fig. 5). The appearance of bistability is clear from Eq. (3), which, by expanding the denominator near the resonance frequency, has the same structure as the nonlinear ferromagnetic resonance curve, demonstrating the foldover effect [44, 45]. In the bistability range, the exact shape of the transmission curve depends on the experimental (simulation) conditions. The solid curves are obtained, when all the simulations start from the same ground state. To access the dashed curve in the simulations, we gradually decrease the excitation frequency starting outside of the bistability range with a constant spin-wave amplitude. The small discrepancy between theory and simulation is mainly attributed to the nonlinear shift coefficient which is extracted from a straight waveguide and not from the ring structure and the previously mentioned nonuniform spin-wave profile in the ring structure.

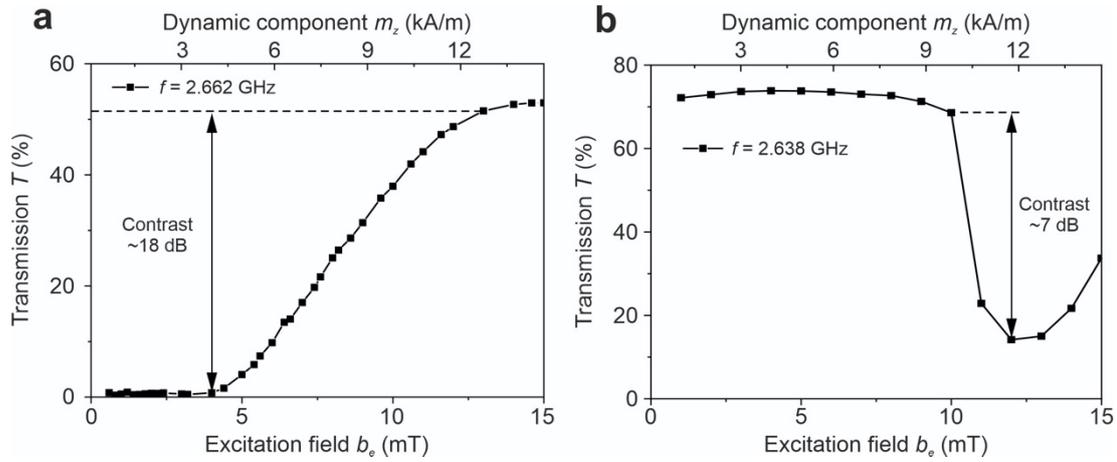

Fig. 6 Simulated transmission T at the fixed frequency of (a) $f = 2.662$ GHz representing an "activation function" and (b) $f = 2.638$ GHz representing a "passivation function" (power limiter) as a function of excitation field $b_e$ and dynamic component of magnetization $m_z$, which linearly scales with $b_e$ in the present field range.

**Nonlinear magnonic ring resonator for neuromorphic applications**

In neuromorphic systems, the so-called "activation function" plays an important role. It describes how an incoming stimulus (here, the incoming spin-wave amplitude $a_1$) is transformed into the output signal (the outgoing spin-wave amplitude $b_1$) via a strongly nonlinear function. Thus, if the ring resonator should serve as a building block for neuromorphic spin-wave computing in a larger network composed of many resonators and interconnecting spin-wave waveguides and combiners, the activation function is characterized by the transmission $T$ of the ring resonator, that should depend strongly on the input amplitude. For instance, in so-called spiking neurosynaptic networks, the "firing" of an artificial neuron, i.e. the emission of an output signal, takes place only if the input signal overcomes a certain threshold. This essentially means that the transmission $T$ should be significantly large only if a certain incoming spin-wave input amplitude is overcome. Wave-based neuromorphic computing using such kind of function has been recently demonstrated for a full neuronal network in optics [30].

In order to characterize the activation function of the magnonic ring resonator, the spin-wave excitation frequency is fixed to $f = 2.662$ GHz, which coincides with the 16[th] resonance in the linear regime, as depicted in Fig. 5 by the vertical dashed line. The output spin-wave intensity nonlinearly depends on the input power because this frequency does not correspond anymore to a resonance in the nonlinear regime.

Figure 6a shows the relative transmission power for $f = 2.662$ GHz as a function of the excitation field amplitude $b_e$ and the dynamic out-of-plane component of magnetization $m_z$ (top axis). The transmission $T$ is almost constant and below 1 % in the excitation field range from 0.6 mT to 4 mT and then strongly increases from $T = 0.78$ % at $b_e = 4$ mT to $T = 51.5$ % at $b_e = 13$mT due to the strong nonlinear shift and the steep slope of the transmission curve versus frequency (compare to Fig. 5). A high contrast of around 18 dB between the output states is observed. This is reflected in the fact that by increasing the input spin-wave intensity by a factor of about 10, the transmitted intensity (absolute spin-wave output power) is increased by a factor of around 700. This strong nonlinearity can be achieved since energy is stored in the ring at resonance. The contrast, threshold and maximum transmission level can be tuned by adjusting the radius of the ring $R$, the transmission coefficient $\tau$, and the loss coefficient $\beta$. As a further example for the use of the ring resonator for data processing, Fig. 6b shows a functionality that can be considered as a kind of "passivation function", meaning that the transmitted intensity decreases with increasing input intensity. Due to the foldover effect shown in the transmission curve in the high-power region (see the magenta line in Fig. 5), the transmission $T$ at the frequency of $f = 2.638$ GHz drops down from 68.6 % to 14.2 % by slightly increasing the excitation field from 10 mT to 12 mT. This functionality can be used to filter out the high-power spin waves or normalize the output spin-wave power, i.e., the absolute spin-wave output power could be made independent of the input spin-wave power in a certain power range. It is worth to note that the precession angle is only around 5 degrees even for the high excitation field of 13 mT which reveals the fact that the energy consumption is very low in the magnon domain.

## Conclusion

In conclusion, a nanoscale nonlinear magnonic ring resonator is proposed, and its functionality is demonstrated using micromagnetic simulations. The transmission curve of the ring resonator in the linear region is of a notch filter type due to the resonant critical coupling effect. Spin waves at resonance frequencies are stored in the ring and cannot pass through it, while spin waves of a frequency in between the resonances pass the ring resonator with only a small loss. Importantly, the nonlinear shift of the spin-wave resonance frequency and, consequently, of the spin-wave phase accumulation, leads to a strong power dependence of the magnonic ring transmission curves. In this nonlinear regime, the resonance frequencies are shifted, the transmission curves become asymmetric and, at large enough input power, exhibit a bistability. The transmission at the linear resonance frequency shows a threshold-like behavior: a low input spin-wave power is stored in the ring structure and the ring only generates an output if the input power exceeds a threshold. This functionality is useful for magnonic logic applications, for instance, in the field of neuromorphic computing. Very different transmission curves can be realized at frequencies not coinciding with the linear resonances. In addition, the transmission functions can be reconfigured by changing the alignment of the magnetization in the ring and the adjacent waveguide. The obtained results are supported by the developed analytical theory, which allows to calculate the ring resonator characteristics in both the linear and nonlinear regimes.

**Methods**

**Spin-wave dispersion in a ring.** The dispersion of spin waves in a magnonic ring can be calculated similarly to those of a vortex-state magnetic disk [46,47]. Note that the ring dispersion relation in the considered system is continuous, $\omega = \omega_k$, and spin waves with continuous wavenumbers $k$ can propagate in it, since the ring resonator is not isolated. In our case, the width of the ring is sufficiently small [40] leading to an almost uniform (unpinned) spin-wave profile across the ring width, which greatly simplifies the calculations. In this approximation, the dispersion relation is given by

$$\omega_k^2 = \omega_M^2 (\lambda^2 k^2 + F_k^{(rr)})(\lambda^2 (k^2 - R^{-2}) + F_k^{(zz)}), \tag{A1}$$

where $\omega_M = \gamma\mu_0 M_s$, $M_s$ is the saturation magnetization, $\gamma$ is the gyromagnetic ratio, $\lambda$ is the exchange length, and $\hat{F}_k = \iint \hat{G}(\boldsymbol{r},\boldsymbol{r}')\exp[ikR(\phi-\phi')]d\boldsymbol{r}d\boldsymbol{r}'/(2\pi Rw)$ is the effective dynamic demagnetization tensor with $\hat{G}(\boldsymbol{r},\boldsymbol{r}')$ being the magnetostatic Green's function in the polar coordinate system [48], and the integration going over the ring surface. For an arbitrary wavenumber, the calculation of $\hat{F}_k$ is complicated. However, for $k_n = n/R$, which are the wavenumbers corresponding the resonant modes of an isolated ring, it is greatly simplified, and yields

$$F_n^{(rr)} = \frac{1}{4Rw}\int_0^\infty f(kh)[I_{n+1}(k) - I_{n-1}(k)]^2 k\, dk, \tag{A2}$$

$$F_n^{(zz)} = \frac{1}{hRw}\int_0^\infty [1-e^{-kh}]I_n^2(k)dk, \tag{A3}$$

where $f(kh) = 1-(1-\exp(-kh))/(kh)$, and we use the notation

$$I_n(k) = \int_{R-w/2}^{R+w/2} J_n(kr)rdr, \tag{A4}$$

with the Bessel functions $J_n$. The function $I_n(k)$ can be expressed via a combination of hypergeometric functions or calculated numerically. The complete continuous spin-wave dispersion $\omega_k$ can be numerically found by interpolation of the dispersion relations of the ring resonance frequencies $\omega_{k_n}$. The spin-wave group velocity is found via $v_{gr} = d\omega_k/dk$. The spin-wave damping rate is calculated using the following general formalism [49]:

$$\Gamma_k = \alpha_G \omega_M (\lambda^2(2k^2-R^{-2}) + F_k^{(rr)} + F_k^{(zz)})/2. \tag{A5}$$

**Coupling between waveguide and ring.** The dynamics of spin-wave amplitudes $a_1(x)$ and $a_2(x)$ in coupled waveguides is described by the following system of equations [24]:

$$\begin{cases} v_{gr}\dfrac{da_1(x)}{dx} = i\omega_c(x)a_2(x) \\ v_{gr}\dfrac{da_2(x)}{dx} = i\omega_c(x)a_1(x) \end{cases}, \tag{A6}$$

where $\omega_c$ is the coupling frequency, which has the meaning of a splitting between the symmetric and antisymmetric collective modes in the coupled waveguide. The difference in dispersion relations (and, consequently, in $v_{gr}$) in the waveguide and ring leads to only a small (second order) correction and is neglected here. The coupling frequency is given by

$$\omega_c = \omega_M \frac{\Omega^{zz} F_{k_x}^{yy}(d) + \Omega^{yy} F_{k_x}^{zz}(d)}{\omega_k}, \tag{A7}$$

and it depends on the coordinate $x$ via the dependence of the distance between centers of straight and ring waveguides $d(x) = d_0 + (R+\sqrt{R^2-x^2})$ with $d_0 = \delta + w$. The position-dependent angle between the waveguide and ring and, consequently, between their static magnetizations, is not accounted for, since in the region which contributes most to the overall coupling, this angle is negligible. Here and below the tensors $\hat{\Omega}$, $\hat{F}$ and $\hat{N}_k$ are defined as in Ref. [24].

From the solution of (A6) one finds the coupling and transmission coefficients, which enter into Eq. (1):

$$\tau = \cos(2\bar{\omega}_c R / v_{gr}), \quad \kappa = \sin(2\bar{\omega}_c R / v_{gr}), \tag{A8}$$

where $\bar{\omega}_c = (1/2R)\int_{-R}^{R} \omega_c(x)dx$ is the "averaged" coupling frequency. This equation can be used for any shape of the coupling area, for example, if the ring is changed to a polygon. For the ring structure, the calculation of $\bar{\omega}_c$ is greatly simplified (note, that $\hat{F}(d)$ is an integral itself) noting that the coupling frequency decays fast with the separation $d$, so we can use the approximation $d(x) \approx d_0 + x^2/(2R)$ and change the integration limits to $(-\infty, \infty)$. Then,

$$\bar{\omega}_c = \frac{\omega_M}{\omega_k}(\Omega^{zz}\Phi_{k_x}^{yy} + \Omega^{yy}\Phi_{k_x}^{zz}), \tag{A9}$$

where

$$\hat{\Phi} = \frac{1}{\sqrt{2\pi R}}\int_0^\infty \hat{N}_k \cos\left(k_y d_0 + \frac{\pi}{4}\right)\frac{dk_y}{\sqrt{k_y}}. \tag{A10}$$

**Micromagnetic simulations**. The micromagnetic simulations were performed using the software package FastMag developed at the University of California, San Diego [50]. This software uses a finite element method to solve the LLG equation and can use the power of modern Graphics Processing Units (GPUs), which leads to the capability to handle ultra-complex geometries at a high speed [50]. The finite element method is especially useful if non-rectangular systems like the presented ring are simulated. The simulated structure of a magnonic ring resonator is shown in Fig. 1a. The parameters of the nanometer-thick YIG are obtained from the experiment as following [36]: saturation magnetization $M_s = 1.4 \times 10^5$ A/m, exchange constant $A = 3.5$ pJ/m, and Gilbert damping for most of the structure $\alpha = 2 \times 10^{-4}$, except for the ring structure. The Gilbert damping in the ring structure is increased to $2 \times 10^{-3}$ to meet the critical coupling condition, and the damping at the ends of the simulated structure is set to exponentially increase to 0.2 to prevent spin wave reflection. The high damping region can be realized in the experiment by placing another magnetic material or a metal on top of YIG. The averaged cell size was set to $10 \times 10 \times 10$ nm$^3$, which is smaller than the exchange length of YIG (~16 nm) and the studied wavelength (~220 nm). To excite a propagating spin wave, a sinusoidal magnetic field $b = b_e\sin(2\pi ft)$ was applied over an area of 40 nm in length, with a varying oscillation amplitude $b_e$ and microwave frequency $f$. The magnetization $M_z(x,y,t)$ was obtained over a period of 250 ns which is long enough to reach a stable dynamic equilibrium. The spin-wave intensity is calculated by performing a Fourier transform from 200 ns to 250 ns, which is long enough to resemble the condition of a dynamic equilibrium. The transmission $T$ is defined by calculating the ratio of the spin-wave intensities in the straight waveguide behind and in front of the ring.

**Data Availability**

The data that support the plots presented in this paper are available from the corresponding authors upon reasonable request.

**Acknowledgement**

The project was funded by the European Research Council (ERC) Starting Grant 678309 MagnonCircuits and the Deutsche Forschungsgemeinschaft (DFG, German Research Foundation) - TRR 173 - 268565370 ("Spin + X", Project B01), the Nachwuchsring of the TU Kaiserslautern. R. V. acknowledges support of National Research Foundation of Ukraine (grant # 2020.02/0261).

**Author Contributions**

Q. W. conceived the idea and designed the structure. P. P. and A. V. C led this project. A. H., V. L., and Q. W. carried out the micromagnetic simulations. R. V. developed the analytical theory. Q. W. wrote the manuscript with the help of all the coauthors. All authors contributed to the scientific discussion and commented on the manuscript.